\documentclass[12pt,preprint]{aastex}
\usepackage{graphicx}
\usepackage{epsfig}
\usepackage{multirow}
\usepackage{footnote}

\shorttitle{IRS Study of BP Piscium}
\shortauthors{C. Melis et al.}

\begin{document}


\title{Shocks and a Giant Planet in the Disk Orbiting BP Piscium?}


\author{C. Melis\altaffilmark{1,2}, C. Gielen\altaffilmark{3,4}, C. H. Chen\altaffilmark{5}, Joseph H. Rhee\altaffilmark{6}, Inseok Song\altaffilmark{6}, B. Zuckerman\altaffilmark{1}}
\email{cmelis@ucsd.edu}


\altaffiltext{1}{Department of Physics and Astronomy, University of California,
Los Angeles, CA 90095-1547, USA}
\altaffiltext{2}{Current address: Center for Astrophysics and Space Sciences, University of California, San Diego, CA 92093-0424, USA}
\altaffiltext{3}{Instituut voor Sterrenkunde, Katholieke Universiteit Leuven, Celestijnenlaan 200D, 3001 Leuven, Belgium}
\altaffiltext{4}{Postdoctoral Fellow of the Fund for Scientific Research, Flanders}
\altaffiltext{5}{Space Telescope Science Institute, 3700 San Martin Dr., Baltimore, MD 21218, USA}
\altaffiltext{6}{Department of Physics and Astronomy, University of Georgia, Athens, GA 30602-2451, USA}


\begin{abstract}
Spitzer IRS spectroscopy supports the interpretation that 
BP Piscium, a gas and dust enshrouded star
residing at high Galactic latitude, is a
first-ascent giant rather than a classical T Tauri star. 
Our analysis suggests that BP Piscium's spectral energy
distribution can be modeled as a disk with a gap that is opened by a 
giant planet. 
Modeling the rich 
mid-infrared emission line spectrum indicates that the solid-state
emitting grains orbiting BP Piscium are primarily 
composed of $\sim$75 K crystalline, magnesium-rich
olivine; $\sim$75 K crystalline, magnesium-rich pyroxene; 
$\sim$200 K amorphous, magnesium-rich
pyroxene; and $\sim$200 K annealed silica (`cristobalite'). 
These dust grains are all sub-micron sized. The giant planet and gap model
also naturally explains the location and mineralogy of the small
dust grains in the disk. Disk shocks that result from disk-planet interaction
generate the highly crystalline dust which is subsequently blown
out of the disk mid-plane and into the disk atmosphere.
\end{abstract}


\keywords{accretion disks --- circumstellar matter --- infrared: stars --- planet-disk interactions --- stars: individual (BP Piscium)}



\section{Introduction}

BP Piscium (BP Psc) is an enigmatic high-Galactic latitude star enshrouded 
in gaseous and dusty material \citep[][hereafter Z08]{zuckerman08a}. Initially,
BP Psc was thought to be an isolated K0-type star 
undergoing its classical T Tauri star phase.
Observations performed by Z08 instead found several indications that 
BP Psc is likely a first-ascent giant star that accretes material from a surrounding
disk of gas and dust while launching bipolar jets. The first such sign that BP 
Psc might not be young was a lithium $\lambda$6708 absorption line
equivalent width that
was $\sim$7 times less than what would be expected for an early
K-type classical T Tauri star. No known example exists in the
literature of such strong lithium depletion for a Solar-like pre-main sequence star.
A second indication that BP Psc was not young came from a low
measured surface gravity. Although young stars typically have lower surface
gravity than main sequence stars (as they are still contracting towards
the main sequence), BP Psc's measured gravity of 10$^{2.5}$ cm s$^{-2}$
was significantly lower than those expected for 1-10 Myr old stars 
\citep[which have gravity values $\gtrsim$10$^{3.5}$ cm s$^{-2}$; see][]{baraffe98} 
and only served to further exacerbate the lithium situation.
Yet another indication, although weaker than the first two, was a discrepancy
between model and kinematically measured masses of BP Psc in the case
that it was a classical T Tauri star. Such problems disappear if instead
it is assumed that BP Psc is a first-ascent giant star.

Although the sum of the observations performed by Z08 
pointed towards BP Psc as an evolved rather than a young star,
they were not necessarily conclusive and further investigation of BP Psc
was warranted. \citet{kastner08} performed molecular line
observations of BP Psc showing that its molecular gas characteristics
are compatible with those found in the expanding envelopes of 
yellow supergiant stars while \citet{kastner10} imaged BP Psc with the 
Chandra X-ray Observatory finding a weak source with characteristics
compatible with rapidly rotating G-type giant stars. 
\citet{melis09} reported the first unambiguous
case of a dusty, accreting first-ascent giant star (TYC 4144 329 2), 
hinting that BP Psc could be part of a previously unidentified class of 
disk-bearing, accreting first-ascent giant stars \citep{melisphd}.
Here we present Spitzer IRS spectroscopic measurements of BP Psc.
These observations were designed to identify
the constituents of BP Psc's dusty disk with the hope that
such additional information would clarify the evolutionary status of
BP Psc and, should BP Psc indeed be a first-ascent giant star,
provide clues as to how it may have been rebirthed
into characteristics typically associated with young, planet-forming stars.


\section{Observations}

BP Psc was observed with the Spitzer Space Telescope Infrared-Spectrograph
\citep[IRS;][]{werner04,houck04} in stare mode on 2007 August 02 with the low-resolution modules 
affording resolving
powers of $\sim$60-120. Since BP Psc is such an infrared bright target, 1 cycle
of 6 seconds ramp duration for each of the two nods was sufficient to obtain well-exposed 
(S/N$\gtrsim$100 per pixel) spectra for each of the Short-Low (SL) and Long-Low (LL)
modules. Data were 
pre-processed with Spitzer Science Center IRS Pipeline 
Version \verb+S16.1.0+ for the SL modules
and \verb+S17.2.0+ for the LL modules. Subsequent bad pixel masking and interpolation
was performed using \verb+IRSCLEAN V1.9+ and a ``master'' bad pixel map. The master
bad pixel map was generated by running the interactive \verb+IRSCLEAN_MASK+
code on each of the four unique basic calibrated data
spectral images per module, then combining these four
output files and the campaign rogue mask that identifies known bad pixels.
The purpose of the master bad pixel map method
was to identify and remove bad pixels that might otherwise
go unnoticed under the well-exposed spectrum.

Individual nods from a pair were subtracted, then these background-subtracted spectra 
were extracted and flux-calibrated using the SPICE software package. Each
extracted, calibrated nod spectrum was compared against its nod pair
to ensure fidelity in the final spectrum. Data from SL order 3 (the so-called
``bonus order'') were removed due to a poor match between the two nods.
Nod-averaged spectral
orders were generally well-matched in terms of flux-continuity. Minor adjustments,
where necessary, were made such that an individual 
order matched the flux of the bluer order. The final spectrum is displayed
in detail in Figure \ref{figirs1} and is overplotted on a complete spectral energy
distribution for BP Psc in Figure \ref{figbppscsed}.

On 2007 July 08 BP Psc was serendipitously imaged at 24 $\mu$m
with MIPS \citep[Multiband Imaging Photometer for Spitzer;][]{rieke04} 
in a program to monitor the dust cloud following
Earth's orbit (AOR 18701312, PI Jayaraman; BP Psc did not fall into the 70 $\mu$m
field of this program). We performed aperture photometry 
on the 24 $\mu$m post-basic calibrated data mosaic image produced
by the Spitzer Science Center MIPS pipeline (version \verb+S16.1.0+).Ê 
We used the Spitzer Science Center recommended aperture correction of 
1.167 for an
aperture radius of 13$^{\prime\prime}$ and sky annulus inner and outer 
radii of 
20$^{\prime\prime}$ and 32 $^{\prime\prime}$, respectively. We detect
a flux density of 5.7 Jy at 24 $\mu$m 
with an uncertainty of 10\%, consistent with the IRAS
25 $\mu$m detected flux and the IRS flux (see Figure \ref{figbppscsed}).




\section{IRS Modeling}
\label{secirsmodel}

Before discussing in detail IRS spectroscopy modeling and results
it is necessary to recall what is known about BP Psc's disk (see Z08 for details). 
Resolved molecular gas emission detected through Submillimeter Array
aperture synthesis imaging indicates that BP Psc has a gas disk in Keplerian
orbit. Keck~II adaptive optics imaging 
resolved a dust disk in scattered light. 
Near- to far-infrared photometric measurements of BP Psc reveal
that $\sim$75\% of BP Psc's observed light is reprocessed into the infrared and that
the orbiting material is well modeled as residing in two regions having 
temperatures of $\sim$1500 and $\sim$210 K (see Figure \ref{figbppscsed}). 
From a Submillimeter Array continuum detection,
Z08 calculate that 7 M$_{\earth}$ of dust orbits BP Psc in the case that
it is a giant star 300 pc distant from Earth. Synthesis of the imaging data
led Z08 to suggest that we view BP Psc's disk close to edge-on
($i$$\sim$75$^{\circ}$) and that the disk has a flared morphology.

\subsection{Disk Structure}
\label{secstruct}

\citet{acke09} and \citet{furlan06} validated the use of 
IRS-based spectral indices to study disk structure by comparing indices computed
from data
to those computed from disk models. We seek to gain insight into BP Psc's
disk structure by comparing its IRS spectral indices to those computed for the disks
studied by \citet{acke09} and \citet{furlan06}. 
We calculate the value of the following indices:
an excess at 7 $\mu$m of 4.55 magnitudes, the ratio of flux at 13.5 $\mu$m to that
at 7 $\mu$m ([13.5/7]) of 0.69 magnitudes, 
and the ratio of flux at 30 $\mu$m to that at 13.5 $\mu$m ([30/13.5]) of 0.96 magnitudes
\citep[see][for how these values are derived]{acke09}; 
the ratio of integrated continuum flux near 6 $\mu$m to that near 13 $\mu$m
(n$_{\rm 6-13}$) of 2.10, the ratio of integrated continuum flux near 13 $\mu$m 
to that near 25 $\mu$m
(n$_{\rm 13-25}$) of 1.93, and the ratio of integrated continuum flux 
near 6 $\mu$m to that near 25 $\mu$m (n$_{\rm 6-25}$) of 2.03
\citep[see][for how these values are derived]{furlan06}.
These values suggest that BP Psc has strong near-infrared excess emission
and steeply rising flux towards longer wavelengths (consistent with
BP Psc's SED; see Figure \ref{figbppscsed} herein and Figure 6 of Z08). 
As mentioned above, the real power of these indices to constrain
the structure of BP Psc's disk comes from comparing them to indices derived
for other well-modeled disk-bearing objects. Such comparisons are presented in
Section \ref{secocd}.

\subsection{Dust Composition}
\label{secbppscirsdust}

The IRS spectrum of BP Psc shows several broad emission lines from solid
state transitions (Figures \ref{figirs1}-\ref{figirs}). To model these features, 
and hence deduce the nature of the emitting particles, we proceed in 
fitting the spectrum via two paths. In one path we attempt to identify
grain species by subtracting the most obvious contributor and
searching for residuals indicative of additional grain species
\citep[as was done successfully for HD 100546 by][]{malfait98}. During these
subtraction fits we try various grain shapes and sizes 
\citep[using the absorption coefficients of][]{min07} to identify the best shape
distribution and grain size to use. In parallel with these subtraction fits, we
fit the data using the $\chi$$^2$ minimization routine
described in \citet{gielen08}. We refer the interested reader
to their work, and references cited therein, for details about model inputs.
The strategy employed by \citet{gielen08} assumes that solid state
emission features originate in an optically thin disk atmosphere. They 
attempt to fit mid-infrared spectra
with a blackbody model for the disk continuum emission and an
emission feature model computed from the linear combination of dust absorption coefficients
multiplied by different blackbody temperatures. 
The $\chi$$^2$ of the model fit to the data is then minimized
by adjusting various parameters; e.g., grain temperature, grain species,
and intensity of grain emission. Both the preliminary $\chi$$^2$ fit and
subtraction fits indicate that the grains producing the emission features
in BP Psc's spectrum are sub-micron ($\sim$0.1-1.0 $\mu$m) in size. 
We also find that both methods obtain consistent 
results in grain species identified.
Similar to what was found by \citet{gielen08}, we find
that dust emission features cannot be reproduced with
spherical grains (e.g., Mie theory), but are instead best 
reproduced with absorption coefficients calculated
using the Gaussian Random-Field (GRF) theory for the grain shape
distribution \citep[][and references therein]{shkuratov05,min07}
and the Discrete Dipole Approximation \citep{draine88}. It is
noted that the grain shape approximations used sometimes
do not reproduce exactly the grain emission features (and hence
likely are not a complete match to the true grain shapes); 
see the discussion in \citet{gielen08}.
In the following three paragraphs we describe the resulting grain 
species identified from these two methods.

Three features $-$ with wavelengths near 33, 28, 
and 23 $\mu$m $-$
stand out in BP Psc's IRS spectrum. These features are
typically associated with crystalline, magnesium-rich olivine grains
(forsterite, Mg$_2$SiO$_4$). 
Fits to the forsterite features left significant residual emission
in the 20-35 $\mu$m range. Experimentation
with various grain species indicates a contribution from 
crystalline, magnesium-rich
pyroxene (hereafter enstatite, MgSiO$_3$). Fits including
enstatite left little residual in the 23-35 $\mu$m range.

Our forsterite and enstatite fits alone could not reproduce the feature
near 10 $\mu$m. The $\sim$10 $\mu$m feature
is typically associated with amorphous dust of the pyroxene or
olivine species. Experimentation with the two suggested that
amorphous, magnesium-rich pyroxene (hereafter pyroxene,
MgSiO$_3$) provided a better fit. 

The remaining residual emission $-$
near wavelengths of 9, 13, 16, and 21 $\mu$m $-$ is 
reminiscent of the features produced by annealed, crystalline silica 
\citep[hereafter cristobalite, SiO$_2$; see][and references therein]{sargent09a}.
It is noted that the 9 and 16 micron complexes are marginally
detected in the residual spectrum. In light of this fact, and in the interest of 
completeness, we reviewed other polymorphs of silica to see if they could 
reproduce the remaining spectral features. $\alpha$-quartz is ruled out
by its characteristic 25 $\mu$m feature which would be quite strong if present
\citep{gervais75}. $\beta$-quartz is ruled out by the high temperatures necessary
to excite its resonances \citep[T$_{\rm dust}$$\gtrsim$850 K;][]{gervais75}; 
our above analysis suggests the small dust grains seen in BP Psc's IRS 
spectrum are significantly cooler (see also below). The polymorphs
coesite and stishovite are also ruled out due to their 
unique emission spectra \citep[see][and references therein]{sargent09a}.
This leaves cristobalite, amorphous silica, and obsidian as potential
silica polymorphs to explain the residual emission. Since obsidian and
amorphous silica have similar spectral features 
\citep[see Figure 3 of][]{sargent09a}, we use amorphous silica to
represent both polymorphs.
To model silica emission features we adopt optical constants 
from \citet{min07} for amorphous silica
(hereafter silica) and from C. Lisse 
\citep[2009, private communication; see the discussion
in the Supplementary On-Line Material to][for a description of the input
optical constants]{lisse06} for cristobalite. Absorption coefficients
are calculated for silica using GRF and for cristobalite
using a continuous distribution of ellipsoids (CDE).
Models using silica result in higher $\chi$$^2$ than do fits using
cristobalite and fits without any SiO$_2$ polymorph provide yet even higher
$\chi$$^2$ values; a comparison of fits using cristobalite, silica, and neither
of cristobalite or silica are shown in Figure \ref{figcrist}. As such, we
consider SiO$_2$ to be significantly detected in the IRS spectrum of
BP Psc and conclude that annealed silica is the most likely polymorph.

Some of the above mentioned transitions are coincident
with those of polycyclic aromatic hydrocarbons (PAHs). PAHs,
however, typically display strong 8.0 $\mu$m features
\citep[see e.g.,][and references therein]{sloan07}. No such
feature is seen in our spectrum of BP Psc, thus we conclude
that PAHs do not significantly contribute to the strength of the 
observed solid state features.

We briefly mention the affect of reddening on mid-infrared solid-state
emission features. Highly reddened stars can exhibit depressed
10 $\mu$m features \citep{gielen08}. BP Psc, at a galacitc latitude of 
$-$57.2$^{\circ}$, is unlikely to be 
experiencing any significant interstellar extinction regardless of its distance.
The dusty material in orbit around this star could in theory provide 
substantial reddening if the grains are small. In their analysis, Z08 
compared low-resolution optical spectra of BP Psc to non-dusty 
stars of similar spectral types and showed that BP Psc is not 
reddened along our line of sight. 
Thus, reddening is unlikely to contribute to the observed strength
of the 10 $\mu$m structure.

With rough grain parameters identified 
we proceed in a final $\chi$$^2$ minimization fit to the spectrum. We restrict
input dust species, grain shapes, and grain sizes to those as determined from
the subtraction and initial $\chi$$^2$ fits. We allow as free parameters
three independent dust continuum temperatures, three independent dust
species temperatures, 
two dust grain sizes ($\sim$0.1-1.0 $\mu$m and $\sim$2.0-4.0 $\mu$m),
and five dust species $-$ forsterite, enstatite, pyroxene, amorphous olivine, 
and cristobalite. We note that the final $\chi$$^2$ minimization routine
employed herein differs slightly from that used by \citet{gielen08}. Our
routine assigns grain species into one of three distinct groups where each group
is allowed its own dust temperature and temperature step size in the model
grid. Amorphous materials (pyroxene and olivine) form one group which has
a temperature grid step of 25 K, crystalline olivine and pyroxene form another
group which has a grid step of 25 K, and silica is in its own group that has
step size of 50 K.
The final model fit is displayed in Figure \ref{figirs}.
Temperatures, mass fractions, associated errors, and an
estimated mass per species are reported in Table \ref{tabbpdust}. Errors
on these parameters are calculated as in \citet{gielen08}: 
we randomly add gaussian noise with a sigma distribution determined
from the spectrum statistical uncertainty per pixel
at each wavelength point. In this manner we generate 100 synthetic spectra,
all consistent with our data, on which we perform the same fitting
procedure. The resulting slightly different fit parameters are
used to derive the mean (our best-fit value) and the standard deviations.
The uncertainty on the mass absorption coefficients are not taken into
account in this error determination. 

Here we summarize some of the main results of the spectral
fitting. No larger sized ($\sim$2.0-4.0 $\mu$m) grains are required in the fit. 
Amorphous olivine (Mg$_2$SiO$_4$, hereafter olivine)
is not necessary to obtain the minimum $\chi$$^2$ model fit.
SiO$_2$, however, is required despite its low mass fraction and its inclusion
in the fit results in a significantly reduced final $\chi$$^2$ \citep[$\sim$84 if
included and $\sim$110 if not; see Section 7 of][for how
we calculate $\chi$$^2$.]{gielen08}
From Table \ref{tabbpdust} it is
clear that the crystallinity fraction of the small dust grains is 
high, $\sim$96\%, with the majority of this coming from forsterite
and enstatite. 

The sub-micron sized dust grains
orbiting BP Psc, in the case that it is a first-ascent giant star, 
would be radiatively blown-out without a spatially 
coincident gas disk (this would not be the case if BP Psc were a T Tauri star). 
To illustrate this we assume BP Psc has a luminosity 
of 100 L$_{\odot}$ and a mass of 1.8 M$_{\odot}$.
With Eq.\ 1 from \citet{chen06}, and assuming grain densities of 3.3 and 2.3
g cm$^{-3}$ for silicates and SiO$_2$ respectively, we estimate that
grains smaller than $\sim$10 $\mu$m in size will be radiatively blown-out of
the BP Psc system in the absence of gaseous material.

\subsubsection{Unmodeled Features}

\begin{figure}
 \begin{center}
 \begin{minipage}[!h]{80mm}
  \includegraphics[width=79mm]{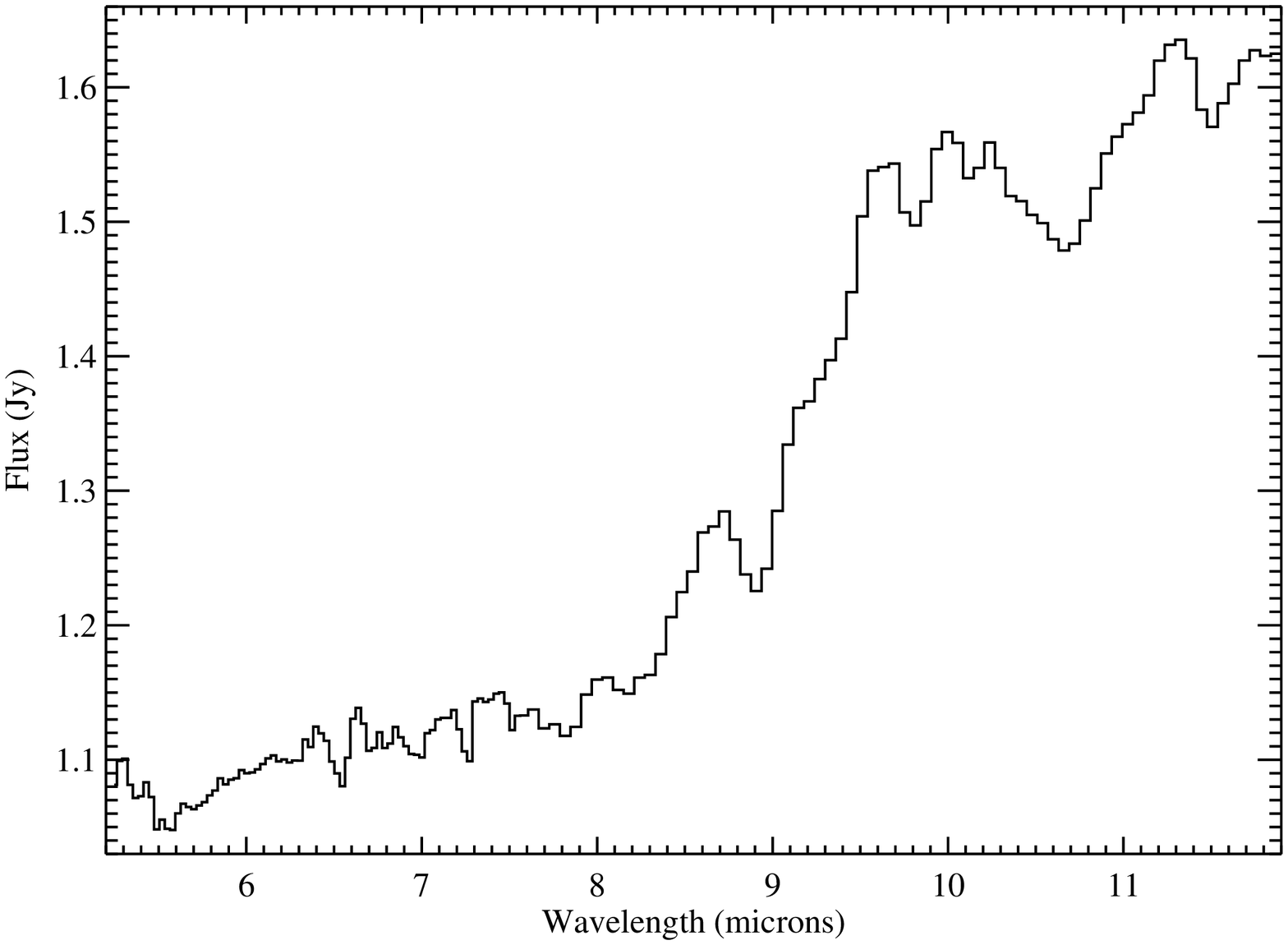}
 \end{minipage} \\*[2.0mm]
 \begin{minipage}[!h]{80mm}
  \includegraphics[width=79mm]{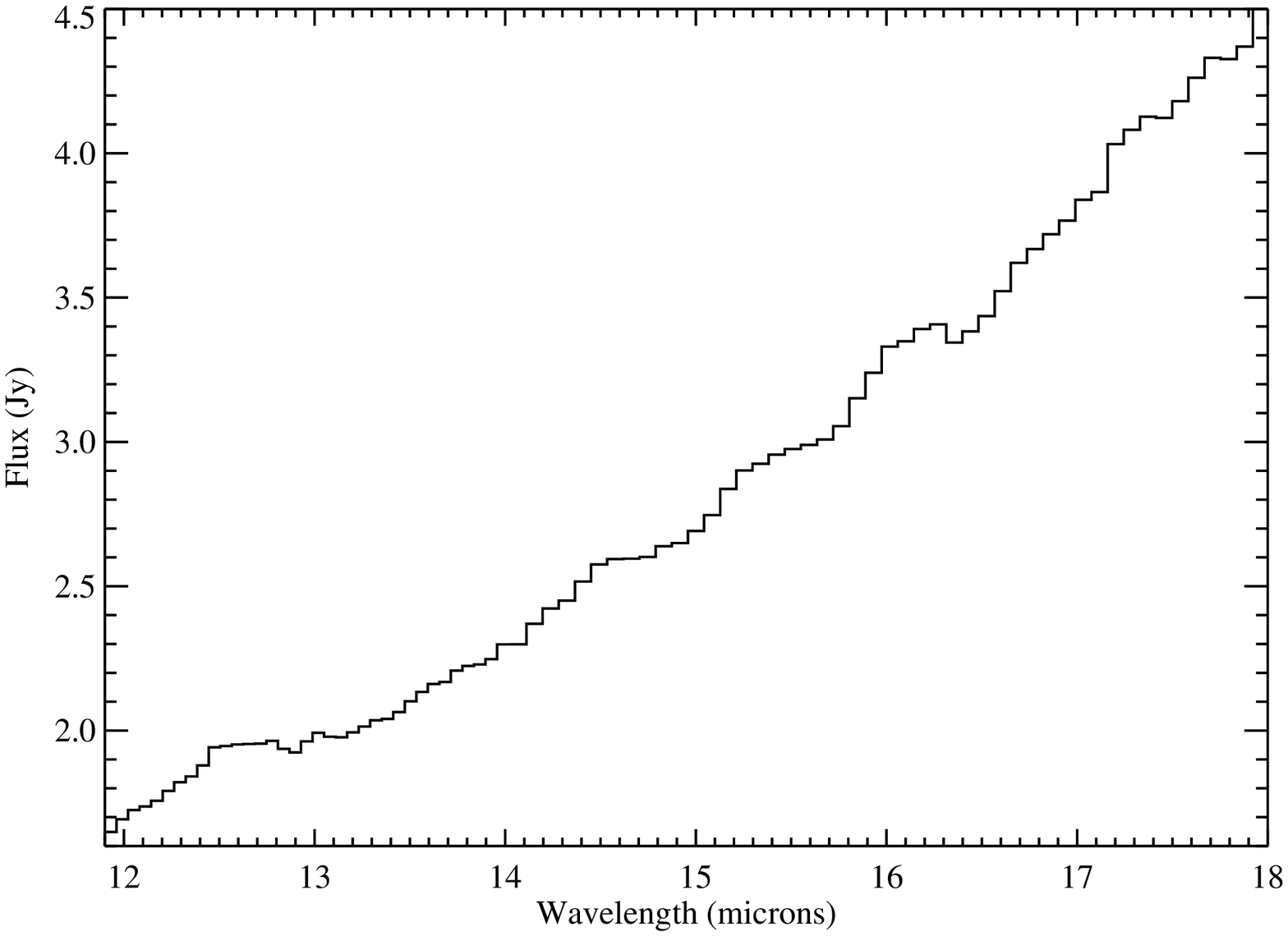}
 \end{minipage}\\*[2.0mm]
 \begin{minipage}[!h]{80mm}
  \includegraphics[width=79mm]{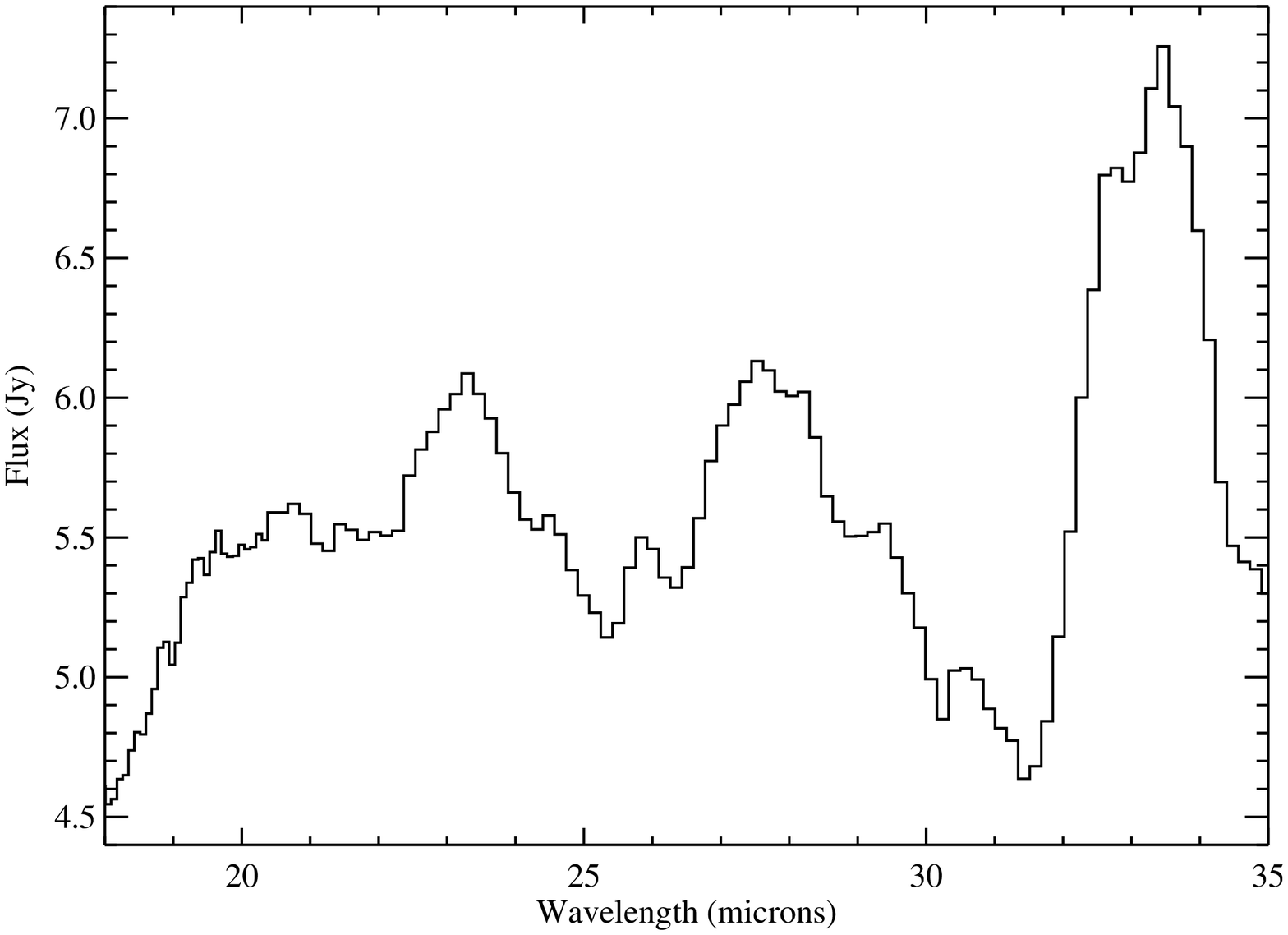}
 \end{minipage}
 \end{center}
\caption{\label{figirs1} {\it All Panels:} Reduced, extracted, and calibrated IRS spectrum of 
               BP Psc. The spectrum S/N per pixel is $\gtrsim$100. 
               {\it Top Panel:} The $\sim$5-12 $\mu$m wavelength range. Weak features in 
               the $\sim$7-8 $\mu$m
               range are dubious as this is an order-overlap region. Note, however, that the
               $\sim$6.5 $\mu$m absorption/emission complex is real and is confirmed in 
               the spectra of both nods. {\it Middle Panel:} The $\sim$12-18 $\mu$m wavelength range.
               Weak features in the $\sim$14-15.2 $\mu$m
               range are dubious as this is an order-overlap region. {\it Bottom Panel:} The 
               $\sim$18-35 $\mu$m wavelength range. Weak features in the $\sim$19-20 $\mu$m
               range are dubious as this is an order-overlap region. Data beyond 35 $\mu$m are
               excluded
               due to the numerous bad pixels that reside in that region of the LL chip.}
\end{figure}

Identification of the $\sim$6.5 $\mu$m absorption/emission complex 
in BP Psc's IRS spectrum
(see Figure \ref{figirs1}) is complicated by its weak signature. It
is possible that this feature arises from the $\nu$$_2$ 
vibrational band of water \citep[see e.g.,][]{woodward07}. This
particular fundamental of gas-phase water
presents rovibrational emission features near 6.5 $\mu$m similar 
in morphology to the weak
features observed at that location in BP Psc's IRS spectrum
\citep[compare Figure \ref{figirs1} herein and Figure 2 of][]{woodward07}.
Since this is not a high fidelity identification, we will refrain
from attempting to interpret the presence of water vapor in BP Psc's
disk. High-spectral resolution, high signal-to-noise spectra
around 6 $\mu$m will be necessary to reliably identify and characterize 
this feature.

There is potentially unmodeled solid-state 
emission remaining near 20 $\mu$m.
If the 20 $\mu$m emitter is a dust species not already present in the model fit, 
then it would be unlikely to have other transitions in the 
$\sim$5-40 $\mu$m range as the rest of BP Psc's IRS spectrum
is reproduced well by the model displayed in Figure \ref{figirs}. 
A potential carrier of the 20 $\mu$m emission feature is FeO nano dust
\citep[][and references therein]{zhang09}. FeO nano dust
hosts one solid-state transition near 20 $\mu$m.
Temperature controls the width and peak wavelength of the FeO
nano dust feature, with temperatures of $\lesssim$100 K shifting the
feature's peak
wavelength to 20.1 $\mu$m and its FWHM to 2.4 $\mu$m 
\citep[][and references therein]{posch04}. Such a peak position
and profile \citep[see also Figure 9 of][]{posch04}
could potentially account for the unmodeled emission in BP Psc's
IRS spectrum.
FeO nano dust can only exist in environments with low ultraviolet
photon density and cool temperatures; ultraviolet photons
will dissociate FeO while high temperatures will
prevent Fe and O from binding together 
\citep[][]{posch04}. Fe and O
orbiting BP Psc at similar locations as the other dust
species would be sufficiently cool to bind together 
while the late spectral type of BP Psc could provide 
the requisite low ultraviolet photon density to allow
the FeO to stay bound.

\section{Comparison to Other Circumstellar Disks}
\label{secocd}

In the following subsections we compare
BP Psc's mid-infrared spectrum to those of 
other well studied disk-bearing objects in an effort
to better understand the physical environment responsible for
BP Psc's disk structure and composition.

\subsection{T Tauri Stars}
\label{secbppscttau}

T Tauri stars are pre-main sequence, Solar-like stars that are still
surrounded by substantial amounts of primordial gaseous and dusty
material. The IRS-indices (see Section \ref{secstruct})
for these disks are consistent with flat,
optically thick, active accretion disks \citep[see e.g.,][]{furlan06}.
BP Psc's IRS-indices are strongly discrepant with those
for the T Tauri stars presented in \citet{furlan06} and \citet{watson09}
$-$ where the Taurus median of 
n$_{\rm 6-13}$, n$_{\rm 13-25}$, and n$_{\rm 6-25}$ are
$-$0.82, $-$0.17, and $-$0.48 respectively $-$
suggesting that the structure of BP Psc's disk 
is significantly different from T Tauri star disk structures
(see Section \ref{secbppschaebe}).

T Tauri star dusty material shows a variety of compositional characteristics
that are at odds or incompatible with what is observed for BP Psc:

\begin{itemize}

\item T Tauri stars sometimes host disks with high crystallinity fractions,
but these grains are typically accompanied by significant quantities of
amorphous species
and are confined to within $\lesssim$10 AU separation from their
host stars \citep{watson09}.
Reasonable age and mass estimates of BP Psc as a classical T
Tauri star (see Table 6 of Z08) would require
the cool forsterite and enstatite grains to reside at separations
$\gtrsim$12 AU.

\item If BP Psc were a young T Tauri star, its disk atmosphere
would have one of the highest
masses in small grains ($\sim$0.1 Lunar masses if located 80 pc from Earth; 
see notes to Table 1) for a star of its class
\citep[e.g.,][quote a range of mass in small dust grains for the T Tauri stars
they studied of $\sim$0.8-700$\times$10$^{-4}$ Lunar masses]{sargent09b,sargent09a}. 

\item \citet{watson09} find in the T Tauri stars they studied 
an anti-correlation between the strength of the 11 $\mu$m 
crystalline olivine feature and the total strength of the entire 10 $\mu$m solid-state
emission complex. Additionally, \citet{watson09} identify a positive correlation between
the total strength of the 10 $\mu$m complex and the magnitude of the
n$_{\rm 13-25}$ index. These correlations suggest that stronger crystalline emission
features, and hence more chemically
evolved grain populations, accompany disk settling
in T Tauri star disks. \citet{sargent09b} found that chemical evolution
doesn't necessarily imply grain growth has occurred, further suggesting that
chemical evolution and sedimentation are tied
\citep[but][show that dust sedimentation alone, without grain growth,
cannot reproduce the observed correlations]{dullemond08}.
BP Psc's grains, despite their nearly 100\% crystallinity fraction, are still 
suspended in the flared disk atmosphere, apparently in contrast with the
observed T Tauri star trends.

\end{itemize}

\subsection{Herbig Ae/Be Stars}
\label{secbppschaebe}

Herbig Ae/Be (HAeBe) stars are the intermediate mass counterparts of classical 
T Tauri
stars. HAeBe stars are much hotter and more luminous 
\citep[median T$_{\rm eff}$$\sim$6000-7000 K, L$\sim$10-1000 L$_{\odot}$
depending on their progress along
pre-main sequence tracks; see][]{palla93} than lower mass, solar-like T
Tauri stars and hence provide a far different circumstellar environment.

BP Psc's spectral energy distribution is reminiscent of Meeus group I HAeBe stars 
\citep[see Figure 4 in][]{meeus01}.
The IRS-indices for BP Psc, when plotted in Figure 2 of \citet{acke09}, lie 
in the region of infrared excess space
describing HAeBe disks having a Meeus group II disk structure
\citep[where group II sources have an outer disk which is protected
against direct stellar radiation by the puffed-up inner disk rim; 
see][and references therein]{dullemond04}. This behavior is peculiar,
but repeated in a well-studied HAeBe star, HD 100546. This source
is given as an exemplar Meeus group I source in \citet{meeus01}.
However, it too resides in the region of Figure 2 of \citet{acke09} describing
Meeus group II disk structures. Recent interferometric work
by \citet{benisty10} has confirmed that HD 100546 has a gap in its
disk (see Section \ref{secplan}). 
It would seem that gapped disks can provide confusing results
when relying on spectral indices alone. Nonetheless, BP Psc's spectral
energy distribution, and hence disk structure, is consistent
with those found for HAeBe stars.

The match of BP Psc's disk structure with those of HAeBe stars suggests that the
inner disk morphology for both types of disk systems is determined
from the same physical effect. It is not yet clear what determines this inner radius
environment, but the ubiquity of hot excess temperatures of $\sim$1500 K that
fit the near-infrared ``bumps'' in HAeBe spectral energy distributions
(thought to arise from puffed-up inner disk rims) 
suggests that dust sublimation is responsible. 
This is in contrast to Solar mass T Tauri stars,
whose inner disk morphologies are determined by magnetospheric
accretion 
\citep[which act on T Tauri disks at $\sim$7 R$_{*}$, close to sublimation temperatures; see][and references therein]{bouvier07}.

ISO, IRS, and ground-based mid-infrared spectroscopy of HAeBe stars
indicates disks sometimes populated by PAHs and/or 
significantly evolved dust as evidenced 
by high mass fractions of larger than sub-micron size, 
chemically and thermally processed 
grains \citep{bouwman01,acke04,vanboekel05,keller08}. 
One HAeBe star, HD 100546,
shows dust emission features in its mid-infrared 
spectrum that are similar to those seen in our IRS spectrum of BP Psc
(Figure \ref{figbppscsed}).
HD 100546 was shown by \citet{malfait98} to have strong
contributions to its mid-infrared spectrum from forsterite and PAH emission.
The grains orbiting HD 100546 are sub-micron sized and originate 
from two distinct temperature regions \citep{malfait98}.
More discussion of HD 100546, and its relation to BP Psc,
can be found in Section \ref{secdoe}.

\subsection{First-Ascent Giant Stars}

Three other first-ascent giant stars have well characterized infrared excess
emission. Two of these objects, HD 233517 \citep[a K2~III star;][]{fekel96,jura03a,jura06} 
and HD 100764 \citep[a $\sim$4400 K, R-type Carbon star;][]{skinner94,sloan07}, 
also have published IRS spectra.
The third, TYC 4144 329 2 \citep[an F2-type star;][]{melis09}, has existing 
IRS spectroscopy that will be mentioned here (a more detailed description
of TYC 4144 329 2's IRS spectrum will appear in 
C.\ Melis {\it et al.}\ 2010 in preparation). 
The infrared excess emission 
for these three objects is modeled as arising from a Keplerian disk of
dusty material that orbits the central giant star. 
In the case of HD 100764 and TYC 4144 329 2,
near-infrared excess emission is present indicating dust as hot as $\sim$1500 K
\citep{skinner94,melis09}.
HD 233517 exhibits infrared excess emission 
indicating a flared, cool dust disk (T$_{\rm dust}$$\sim$70 K)
orbiting at $\sim$45 AU separation from the giant star \citep{jura03a}.

IRS spectroscopy of all three of the above mentioned giant stars indicates that
they are orbited by PAHs and small, amorphous, silicate dust grains 
\citep[][C.\ Melis {\it et al}.\ 2010 in preparation]{jura06,sloan07}. It is interesting
to note that all three of these giant stars host obvious PAH features while
BP Psc does not (see Figure \ref{figirs1}; 
although the $\sim$6.5 $\mu$m feature in BP Psc's IRS
spectrum remains unidentified, it is unlikely to be the
characteristic 6.2 $\mu$m PAH transition).

\subsection{Post-AGB Binaries}

With the IRS and TIMMI2 mid-infrared spectrographs, \citet{gielen08} 
performed a comprehensive study of the circumstellar
material in Keplerian orbits around post-AGB binaries  
having a median stellar effective temperature of
$\sim$6000 K and assumed luminosities in the range of
$\sim$5000$\pm$2000 L$_{\odot}$.
The disks around post-AGB binary systems
are well-modeled as passive, irradiated disks \citep{deroo07a,deroo07b}.
Some of the post-AGB binary IRS spectra presented in \citet{gielen08}
have similar slopes as the BP Psc IRS spectrum, indicating that
they could have similar structure.

\citet{gielen08} found
that the stable disks orbiting post-AGB binaries host populations
of hot and cold amorphous and crystalline grains. Grains in post-AGB binary dust 
disks show evidence for strong processing,
both in the form of grain growth beyond sub-micron sizes and
crystallization.
The physical process by which disks form in post-AGB binary systems is not well 
understood, but it is expected that interactions between the evolved star and 
its companion play a necessary role \citep{deroo07b}. 

Similar to our discussion of radiative blow-out for grains orbiting BP Psc,
it can be shown that post-AGB binaries must also have significant quantities 
of gas spatially coincident with their dusty material. Since the minimum grain size
that can orbit a star scales linearly with luminosity, and since post-AGB binary
systems have similar masses as first-ascent giants,
we estimate that the minimum size grains that can orbit a post-AGB binary
in the absence of gas would be $\gtrsim$100 $\mu$m in size.

\section{Implications for the Evolutionary State of the BP Psc System}

Z08 established through various methods that BP Psc has the effective 
temperature of a G8-K0 star and likely has the luminosity of a first-ascent red 
giant star.
We seek to test the luminosity result under the assumption that the effective 
temperature is correct. 

Comparison of BP Psc's dust mineralogy to those of other classes of
disk-bearing objects does not provide any significant insight into
BP Psc's evolutionary status. Although there are discrepancies
with what is observed for classical T Tauri stars, there are also
discrepancies with the constituents observed in disks
around first-ascent giant stars; for example,
other giant stars exhibit detectable PAH emission whereas
BP Psc does not. 
BP Psc, with an accretion rate of
10$^{-8}$ M$_{\odot}$ yr$^{-1}$ (Z08), almost certainly emits sufficient ultraviolet
radiation to excite PAH emission; especially if late-type giant stars 
like HD 233517 and HD 100764, who do not have any obvious accretion, can excite
PAH grains. The absence of obvious PAH emission suggests that BP Psc lacks
abundant PAH molecules (perhaps the effect of carbon-poor
disk material) and/or that the strong near-infrared continuum 
emission from BP Psc's inner disk rim veils PAH emission features 
\citep[e.g.,][]{geers06}. In terms of the presence of a crystalline dust population
and lack of PAHs, BP Psc's dust characteristics most resemble those of 
post-AGB binaries (it is noted that Z08 demonstrated that BP Psc cannot have
a luminosity as large as that of an AGB star). 
BP Psc's dust differs from the objects
studied by \citet{gielen08} as the majority of BP Psc's grains are sub-micron
sized and cool (T$_{\rm dust}$$<$300 K) while the majority of post-AGB binary stars exhibit 
hot grains (T$_{\rm dust}$$\gtrsim$300 K which results in strong emission features
in the $\sim$10 $\mu$m region)
with sizes larger than $\sim$1 $\mu$m.

As discussed in Sections \ref{secirsmodel} and \ref{secbppschaebe},
BP Psc's disk has striking structural similarities to disks around HAeBe stars; 
in particular, the necessity for the puffed-up inner disk rim that is characteristic 
of HAeBe disks. This result suggests that BP Psc, if a classical T Tauri star, lacks 
strong magnetic fields that effectively funnel material onto the star and prevent 
the build-up of a substantial inner disk rim. It is hard to imagine BP Psc as a 
classical T Tauri star without strong magnetic activity \citep[e.g.,][]{preibisch05}. 
Measuring Zeeman splitting of BP Psc's atmospheric absorption lines could provide 
concrete evidence against the system being pre-main sequence should the lack of 
a strong magnetic field be verified \citep[][favor a scenario where
BP Psc's X-ray emission is generated through either coronal processes
or star-disk interactions, both of which require BP Psc to be magnetically
active]{kastner10}. 

\section{BP Psc's Disk Origin and Evolution}
\label{secdoe}

Analysis of our IRS data suggests that BP Psc is more likely an evolved
star than a T Tauri star, consistent with the results obtained by Z08 and
\citet{kastner08}. How did this giant star come to be orbited by so much
dusty and gaseous material?
Z08 postulate that BP Psc, if
an evolved star, could have been rebirthed into its young star-like dusty and
gaseous environment by consuming a short orbital-period companion at the onset
of its post main-sequence evolution. Such a disk formation mechanism is also 
favored for the dusty first-ascent giants TYC 4144 329 2 
\citep{melis09} and HD 233517 \citep{jura03a}. Unfortunately, the literature 
is lacking binary engulfment models that produce results  
consistent with observed outcomes $-$ giant stars bearing disks 
in excess of their stellar 
radius and no evidence for the hypothetical consumed companion. The model that
provides the most reasonable match to the final observed outcome is outlined
in \citet{shu79}. Their analytical model entails
interaction between an evolving star and its short-orbital period,
less massive secondary that causes
matter to leave the system through the L2 lagrangian point in a spiral pattern.
Eventually the spiral outflow of material will return to strike itself resulting in the 
formation of a stable disk of material that accretes onto the central giant star
and
whose maximum size is governed by conservation of angular momentum.

\subsection{Does BP Psc's Dust Composition Result from Binary Engulfment?}
\label{secbppscbe}

According to the \citet{shu79} model, material lost from binary
interaction will initially be close to the host giant star and thus would
be subjected to the intense giant star radiation field. Once this material
passes beyond the sublimation radius it can begin to coalesce into
dust grains. These dust grains should be efficiently annealed with the heat
from the nearby luminous star, resulting in a high fraction of dust being crystalline.
In their study of the 10 $\mu$m emission feature of HAeBe stars, 
\citet{bouwman01} found that with increasing abundances of forsterite 
one expects to see larger mass fractions of crystalline SiO$_2$.
Such a result is indicative of the thermal annealing process which transforms
amorphous into crystalline grains with SiO$_2$ as a by-product. That we see
SiO$_2$ in the material surrounding BP Psc lends some support to
a disk formation model similar to that presented by \citet{shu79}.
The orbital semi-major axis of this crystalline material will continue to expand
until the disk reaches its maximum size. 
Thus, highly crystalline material would be transported to the
cooler, outer regions of the disk where it is observed.

There are some problems with the above grain evolution scenario. One is that other
first-ascent giants with dusty material thought to originate
in a manner similar to BP Psc's do not show evidence for highly crystalline material.
A second problem is that the polymorph of SiO$_2$ most likely present 
in BP Psc's IRS
spectrum, cristobalite, must be cooled very quickly after being annealed
or it will revert to lower temperature polymorphs that do not exhibit the
characteristic 16 $\mu$m emission feature \citep{sargent09a}. It is difficult
to imagine that the disk excretion could be slow enough to efficiently anneal
SiO$_2$ into a cristobalite polymorph and at the same time rapid enough to transport this
cristobalite to cooler regions before it can revert. More detailed
modeling and simulations of disk formation through engulfment
could further explore these issues.

\subsection{A Planetary Model to Account for BP Psc's Observed Disk Structure and Composition}
\label{secplan}

It seems unlikely that the composition of the dust orbiting BP Psc is linked to
its disk formation mechanism due to the problems discussed
above.
Instead, the dust mineralogy could be the
result of evolutionary processes which occur in protoplanetary disks
and are understood to be the precursors to (or indicators of) 
planet/planetesimal formation. Under such an assumption our IRS results could 
provide insight into whether or not new planetesimal formation is 
occurring in BP Psc's disk. As discussed 
above, 
the presence of cristobalite would require an event that heated the disk to temperatures
of $\sim$1000 K and then allowed it to cool rapidly. Such an event likely took
place in BP Psc's outer disk where the disk equilibrium temperature is 
sufficiently low to prevent the cristobalite from reverting \citep{sargent09a}. As
discussed in \citet{sargent09a}, disk-lightning or nebular shock 
scenarios can explain the presence of cool,
highly crystalline material (especially cristobalite). However, they 
do not explain why so much of
this material would be in BP Psc's flared disk atmosphere \citep[lightning and
shocks are expected to mainly act in the disk interior;][]{pilipp98,harker02},
especially when one considers 
that grain processing should be linked with grain growth
and sedimentation \citep[e.g.,][and references therein]{natta07,watson09}.

\begin{figure}
 \begin{minipage}[!h]{80mm}
  \includegraphics[width=79mm]{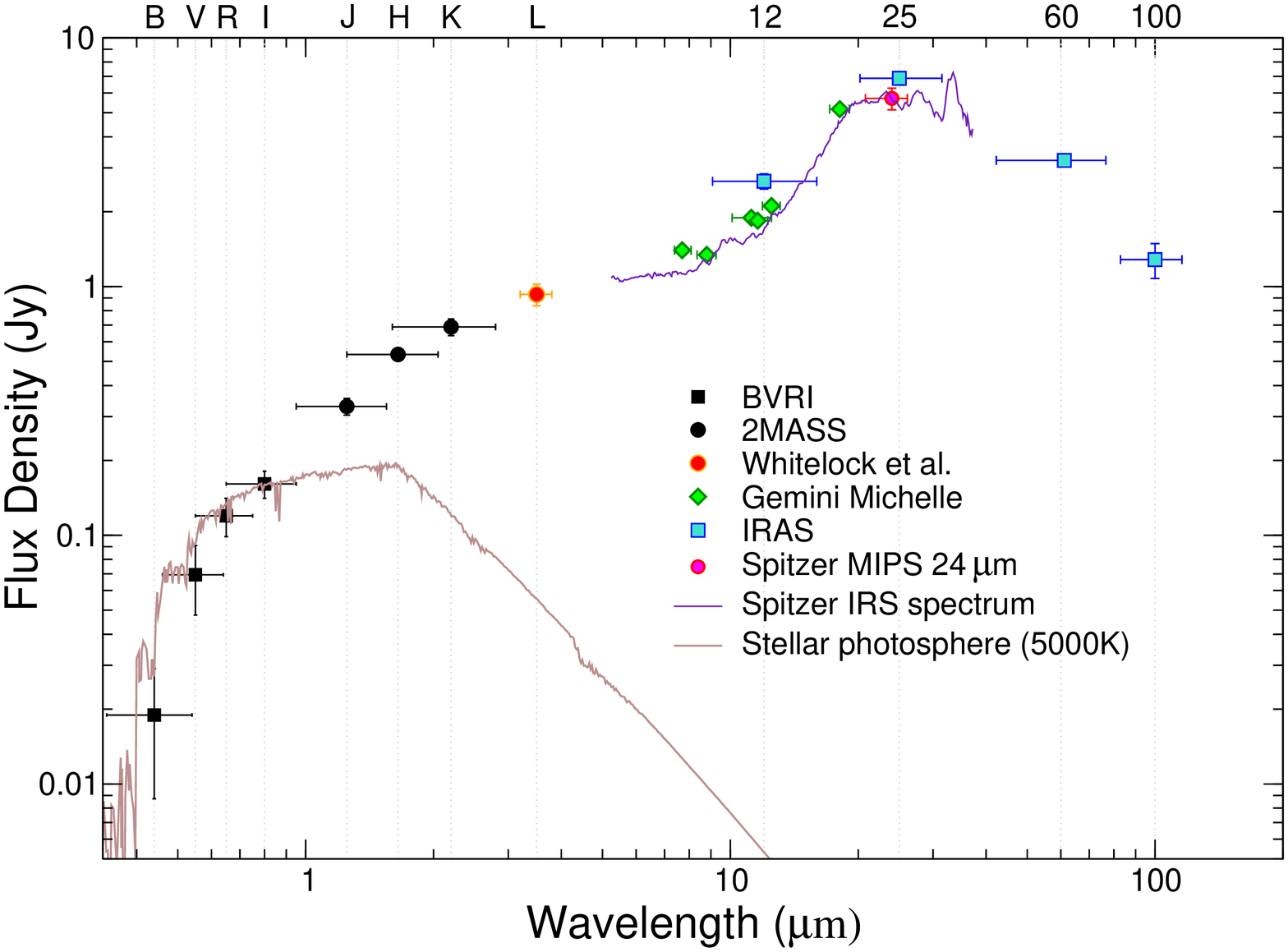}
 \end{minipage}
 \begin{minipage}[!h]{80mm}
  \includegraphics[width=79mm]{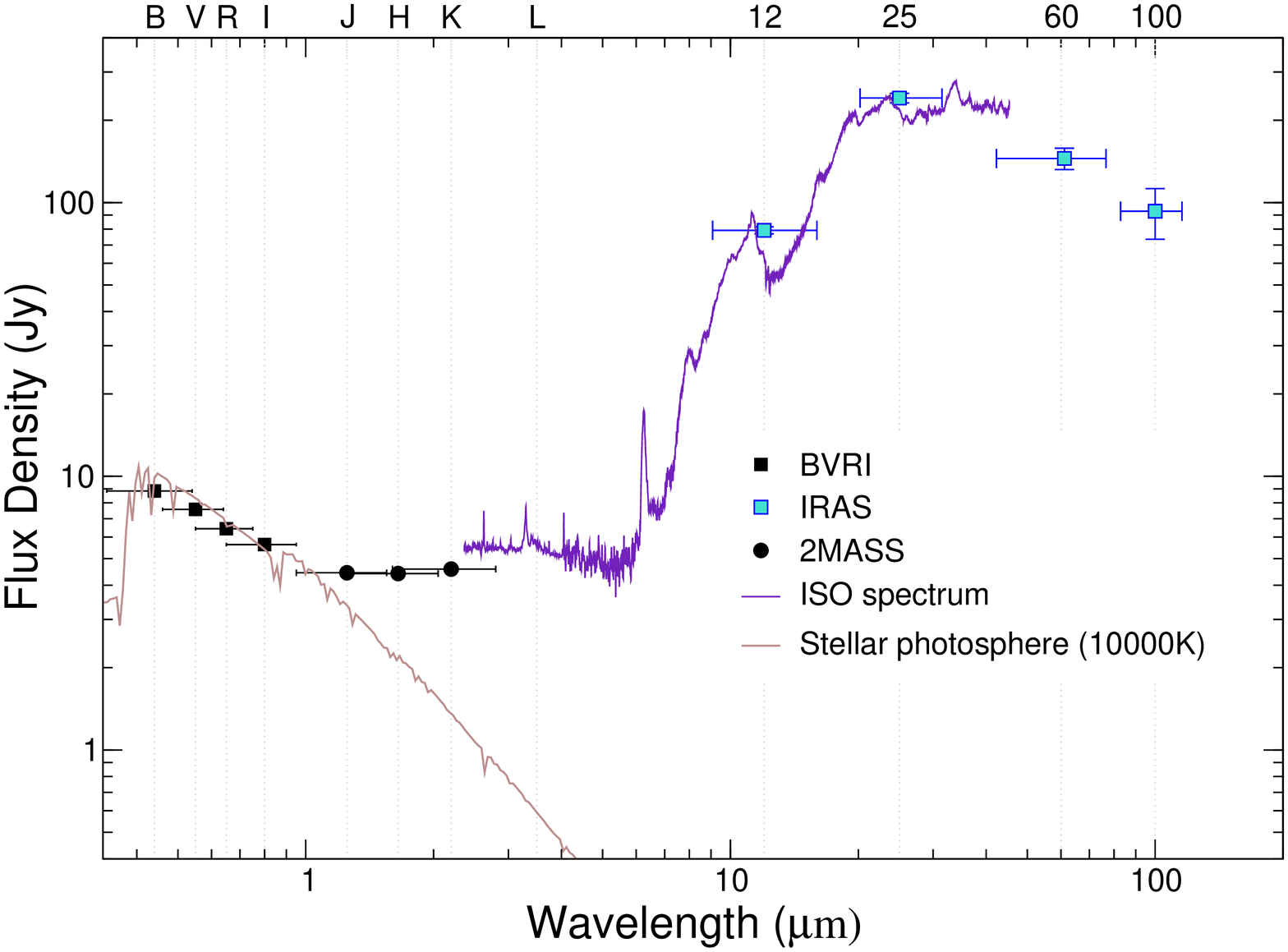}
 \end{minipage} \\*[2.0mm]
 \begin{minipage}[!b]{80mm}
  \includegraphics[width=79mm]{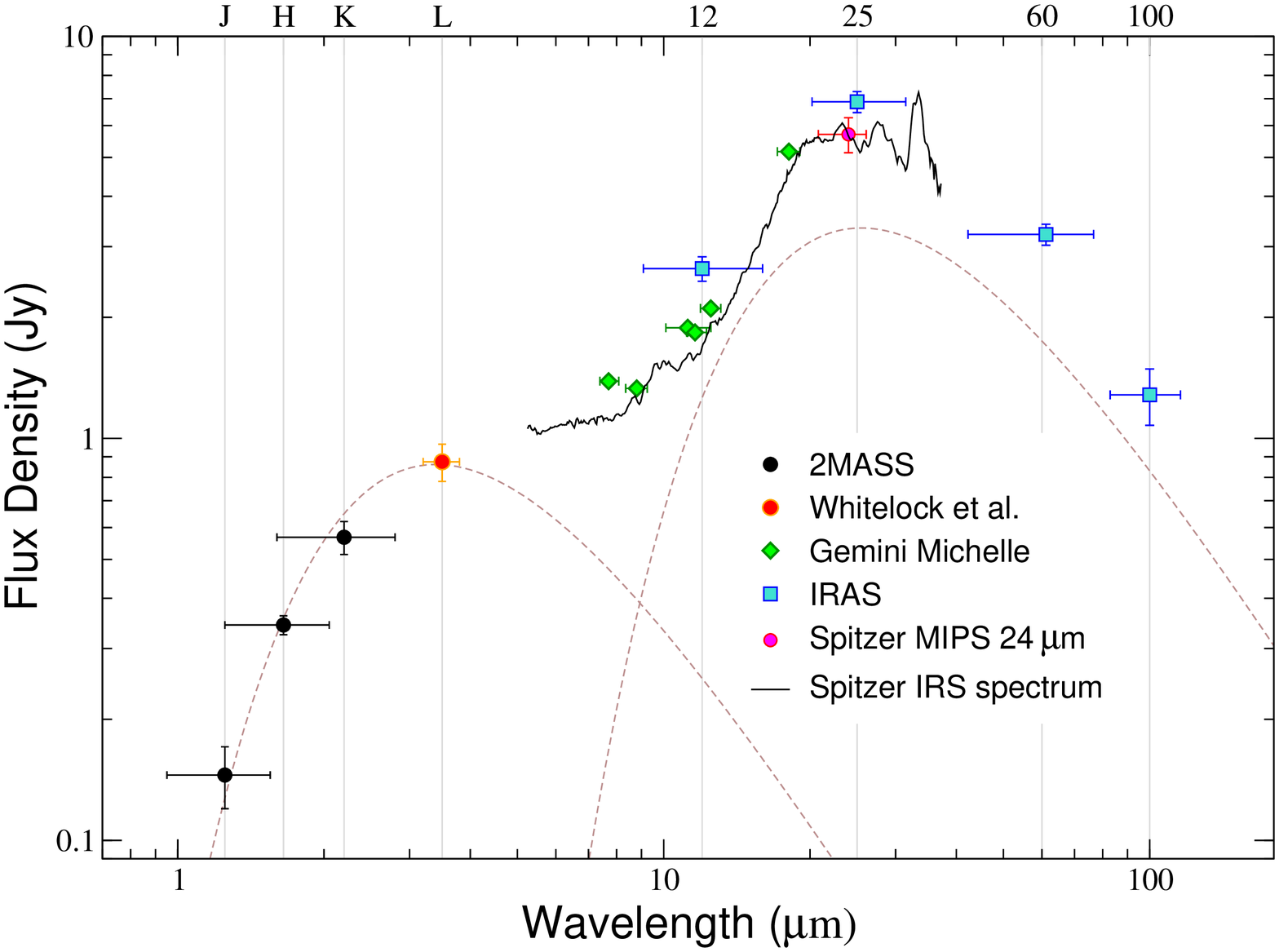}
 \end{minipage}
 \begin{minipage}[!b]{80mm}
  \includegraphics[width=79mm]{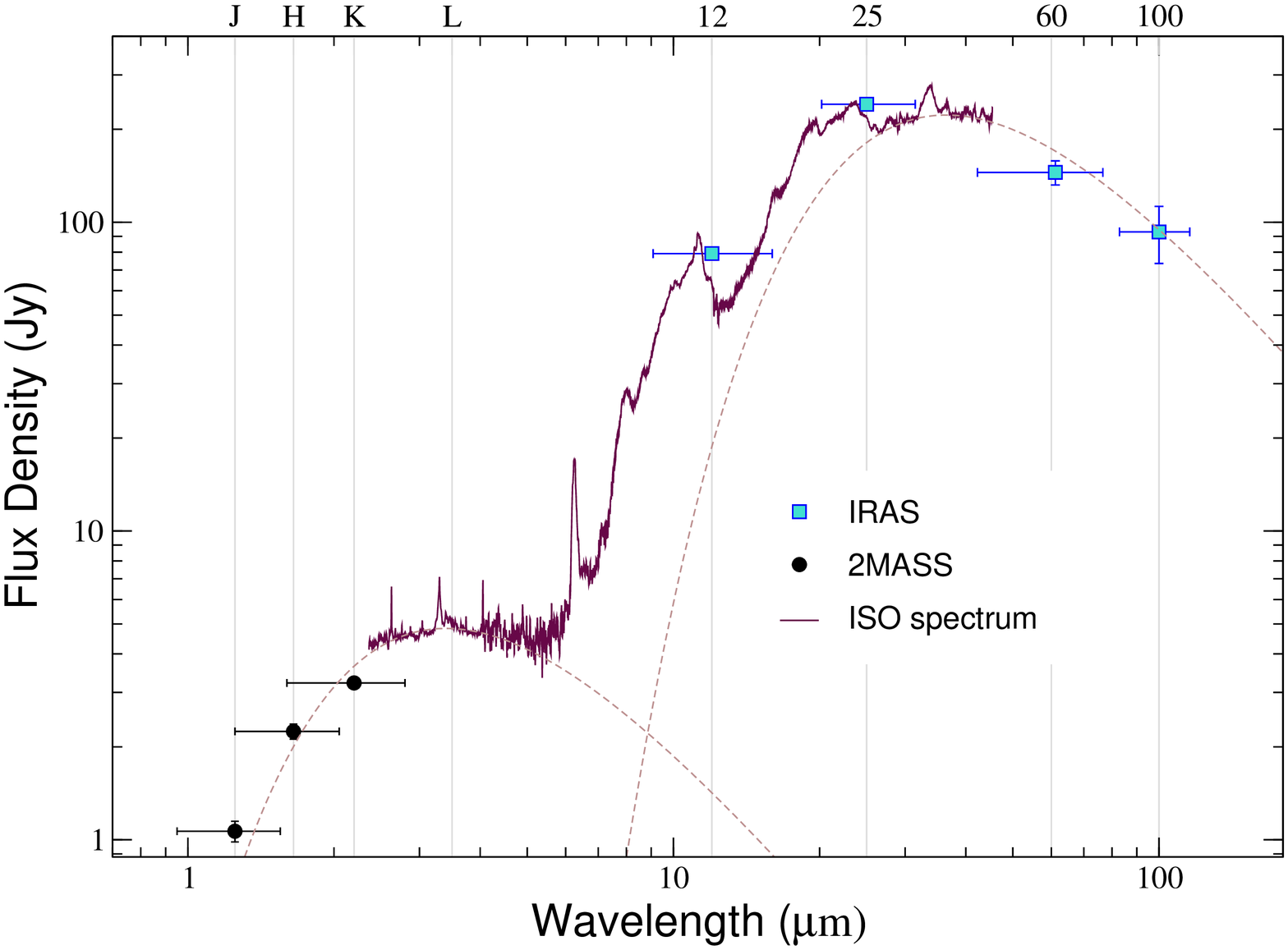}
 \end{minipage}
 \caption{\label{figbppscsed}{\it Top Left Panel:} Spectral energy distribution of 
                BP Psc \citep[from][]{zuckerman08a}
                with the reduced, extracted, and calibrated
                IRS spectrum and MIPS 24 $\mu$m measurement 
                overlaid. {\it Top Right Panel:} Spectral energy distribution of
                HD 100546 with the calibrated ISO-SWS spectrum overlaid \citep{malfait98}.
                A 10000 K stellar photosphere model \citep{hau99} has been fit to the BVRI
                data points.
                {\it Bottom Left Panel:} BP Psc measurements after subtracting the 5000 K photospheric
                model. The residual disk flux is fit with two blackbodies (brown, dashed
                lines) having temperatures of 1500 K and 200 K, respectively. The 200 K blackbody
                displayed is similar to the $\sim$200 K blackbodies used in fitting the dust continuum
                emission in the IRS spectrum (see Figure \ref{figirs}).
                {\it Bottom Right Panel:} HD 100546 measurements after subtracting the 10000 K
                photospheric model. The residual disk flux is fit with two blackbodies (brown, dashed
                lines) having temperatures of 1500 K and 200 K, respectively.}
\end{figure}

\begin{figure}
 \begin{center}
  \includegraphics[width=130mm]{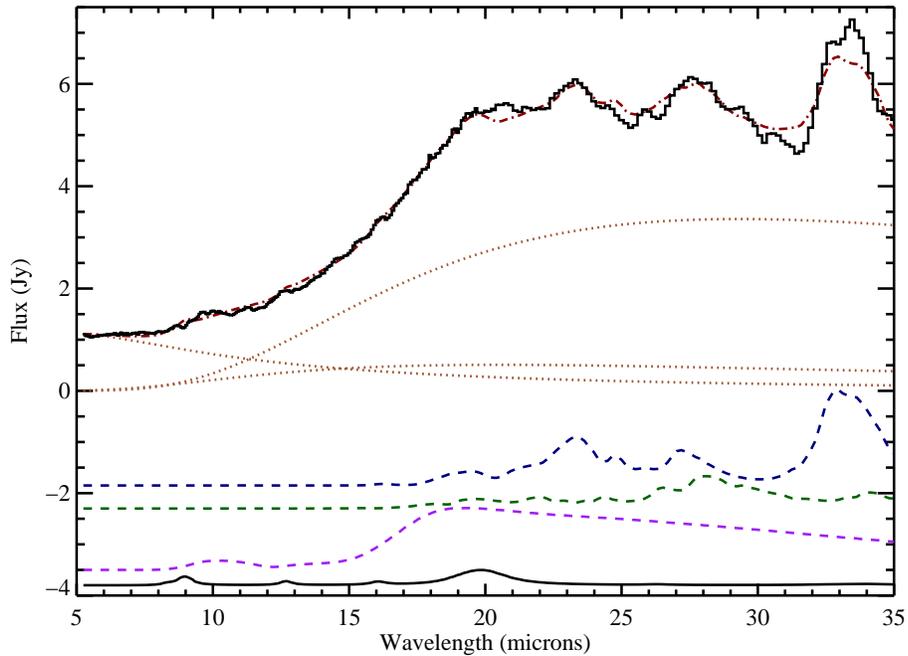}
 \end{center}
\caption{\label{figirs} IRS spectrum of BP Psc; the spectrum S/N per pixel is $\gtrsim$100.
               Model fits to the data (upper black curve) are overplotted with individual
               dust species' emissivities offset by negative flux for clarity. From bottom to top the
               models are for: cristobalite (black solid curve), pyroxene (purple dashed curve),
               enstatite (green dashed curve), forsterite (blue dashed curve), blackbodies
               (three brown dotted curves) having temperatures of 175, 250, and 1000 K respectively,
               and the sum of all the aforementioned curves (red dot-dashed curve).
               See Table \ref{tabbpdust} for the dust parameters.}
\end{figure}

\begin{figure}
 \begin{center}
 \begin{minipage}[!h]{80mm}
  \includegraphics[width=79mm]{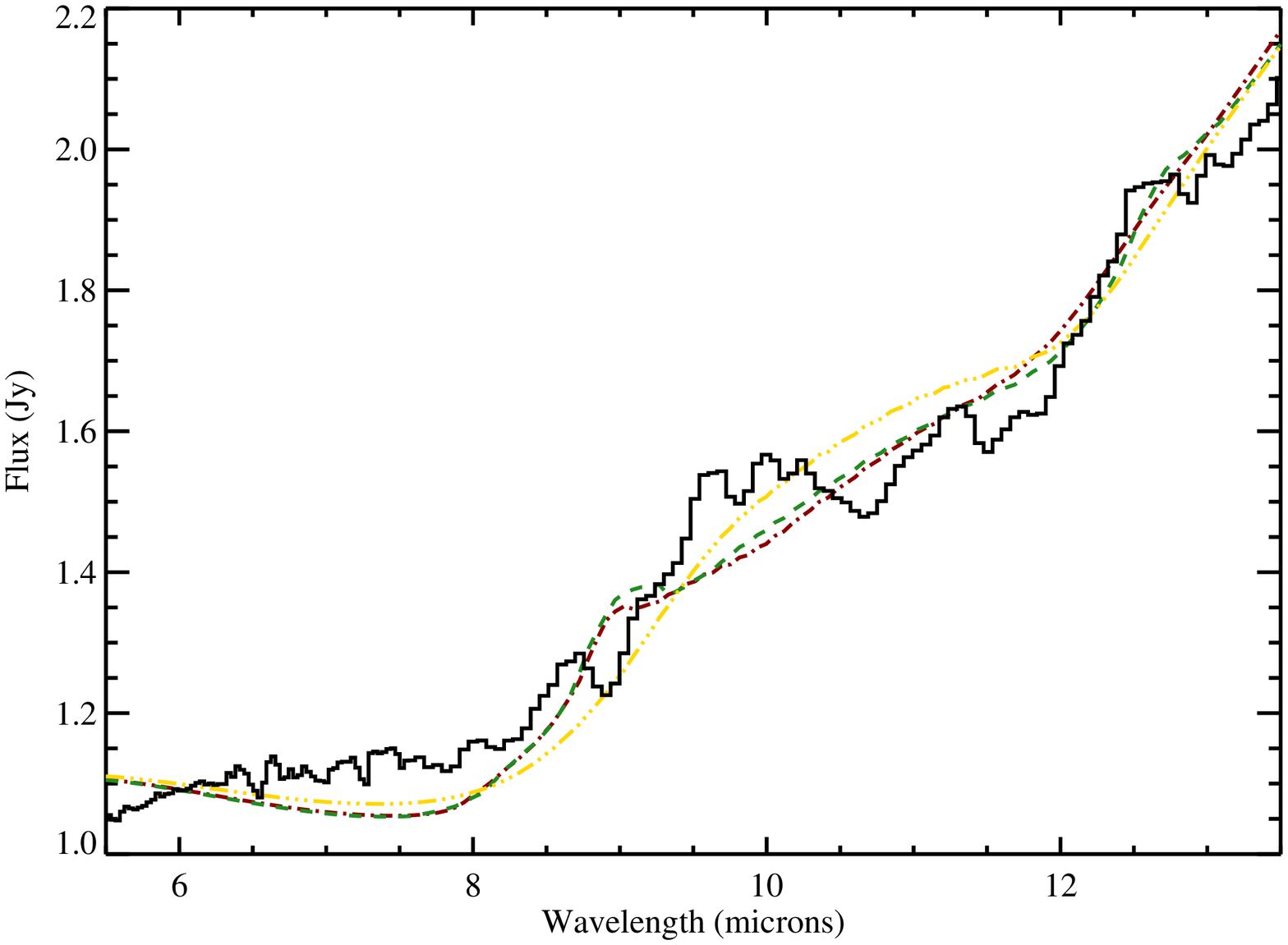}
 \end{minipage} \\*[2.0mm]
 \begin{minipage}[!h]{80mm}
  \includegraphics[width=79mm]{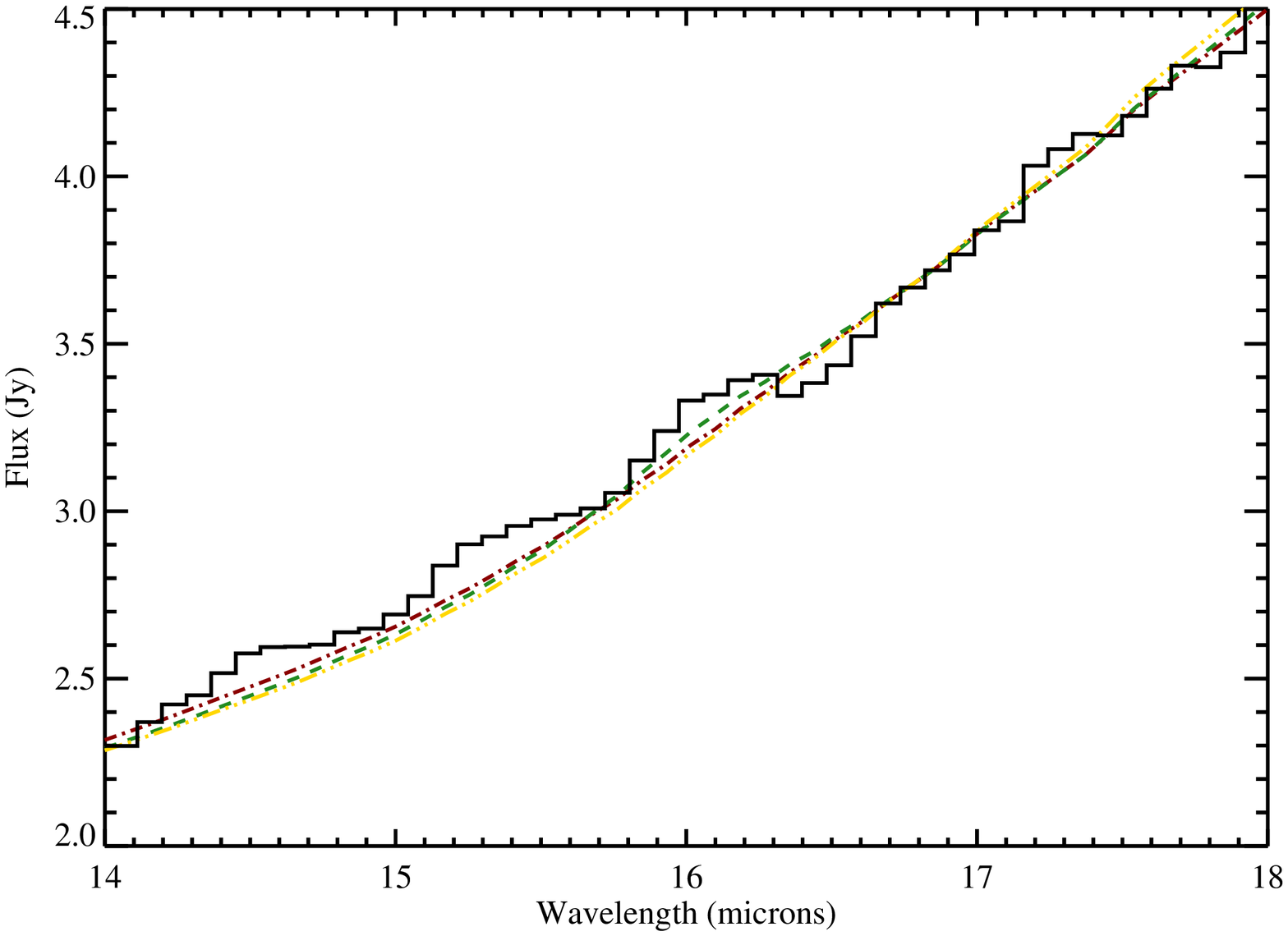}
 \end{minipage}\\*[2.0mm]
 \begin{minipage}[!h]{80mm}
  \includegraphics[width=79mm]{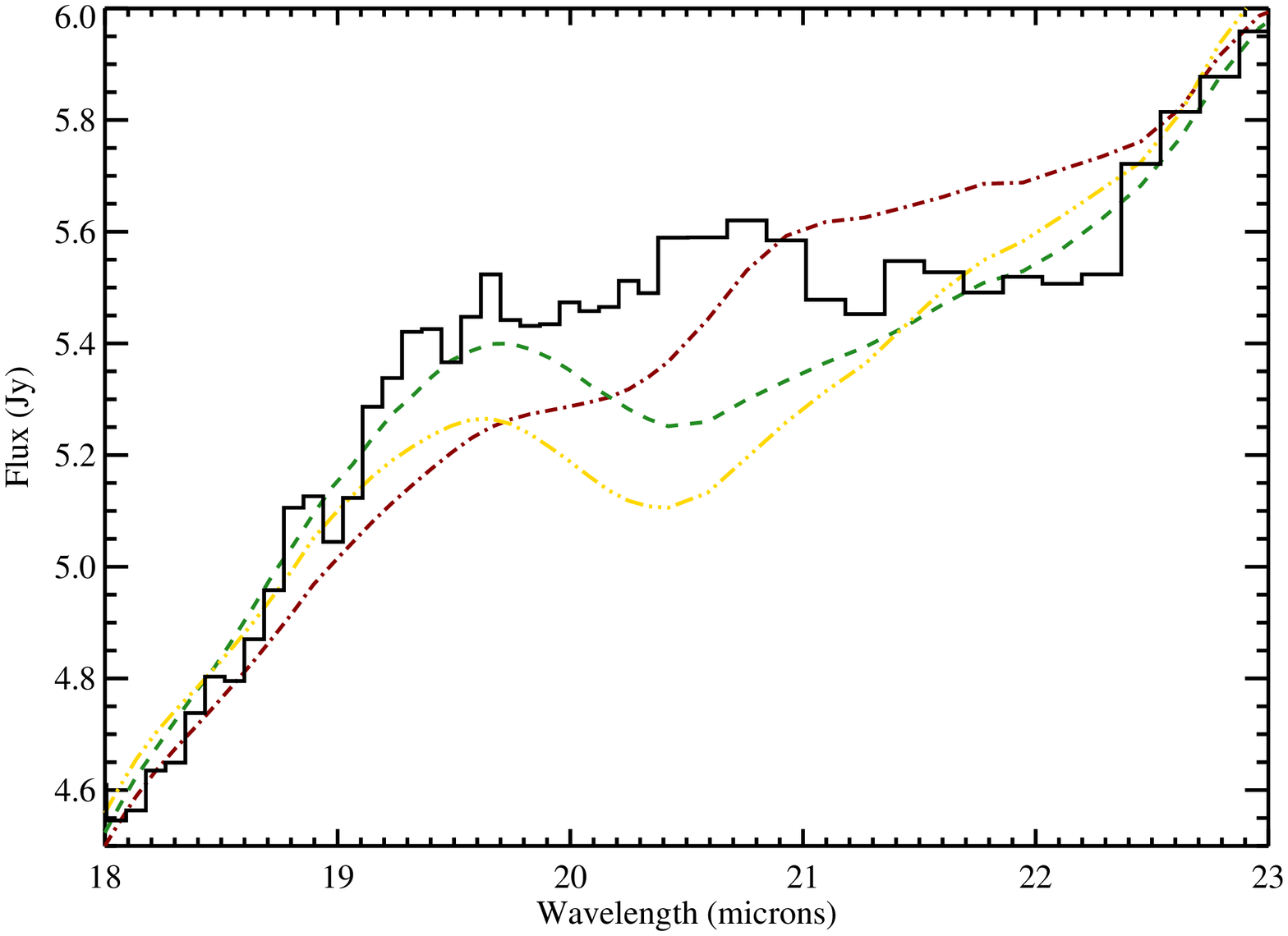}
 \end{minipage}
 \end{center}
 \caption{\label{figcrist} Regions of BP Psc's IRS spectrum having features from SiO$_2$;
                the spectrum S/N per pixel is $\gtrsim$100. 
                Overplotted are final dust models using cristobalite (green, dashed line; see
                also Figure \ref{figirs}), silica (red, dash-dot line $-$ see Section
                \ref{secbppscirsdust}), and a model with no silica or cristobalite 
                (gold, dash-dot-dot-dot line). Fits using cristobalite result in
                lower $\chi$$^2$ values than those
                employing silica and no silica or cristobalite. Cristobalite performs better
                around 12.7 $\mu$m, 16 $\mu$m, and 20 $\mu$m.}
\end{figure}

Solutions to these problems come from comparing BP Psc to
HD 100546. \citet{bouwman03} suggest that HD 100546, who presents
a spectral energy distribution (Figure \ref{figbppscsed}) and crystalline silicate emission 
similar to those of BP Psc (with the exception of the presence of
SiO$_2$), has a proto-gas giant planet
carving a gap within its disk. Such a gap naturally explains the enhanced mid-
to far-infrared excess emission relative to the near-infrared 
excess emission in HD 100546 when compared to other HAeBe stars as
the far side of the gap would be directly irradiated by the star. This
direct stellar irradiation would produce a vertically extended wall in the disk,
allowing it to intercept a large fraction of stellar light which would then
be reradiated as blackbody emission having the characteristic temperature
for the disk gap location ($\sim$200 K for HD 100546 and BP Psc;
see Figure \ref{figbppscsed}). Interferometric observations of HD 100546
by \citet{benisty10} confirm that HD 100546 does indeed have a gap
within its disk. Recent optical 
and ultraviolet spectroscopic studies of HD 100546 further support the 
interpretation that this star has a giant planet orbiting within the gap in 
its disk \citep{grady05,acke06}. Since BP Psc and HD 100546 have
similar SEDs (Figure \ref{figbppscsed}) and hence similar disk structure, 
it is reasonable to surmise that the same
mechanism responsible for HD 100546's disk structure operates also 
in BP Psc's disk. Thus, it is plausible that a giant planet orbits and opens
up a gap within BP Psc's disk. Such a giant planet would have an orbital
semi-major axis of $\sim$4 AU (assuming BP Psc has a luminosity of 
$\sim$100 L$_{\odot}$).
We note that theoretical works predict that any such
object have a mass that is $<$10 M$_{\rm Jup}$ as more
massive bodies would restrict the flow of material towards the star
resulting in a depleted inner disk \citep{lubow99}.
Such depletion is 
not compatible with observations that indicate large quantities
of gaseous and dusty material in BP Psc's inner disk (Z08).
It is not clear whether this giant planet would be from BP Psc's original planetary
system or if it would be a protoplanet forming from the material in 
BP Psc's giant star disk.

Two mechanisms to explain the observed dust
mineralogy are discussed in the following subsections. 
Both rely on the existence of a giant
planet opening a gap within BP Psc's disk.

\subsubsection{Colliding Planetesimals}

The giant planet orbiting within BP Psc's disk could 
stir up rocky bodies that also orbit in the disk gap. These rocky
bodies could be remnant planetary building material left over
from BP Psc's first planetary system that orbit at similar stellar
separations as the giant planet (like Jupiter's trojan asteroids)
or they could have formed in-situ in BP Psc's revived giant star disk.
Following \citet{bouwman03}, we expect ensuing collisions between these
stirred planetesimals to release previously generated crystalline material. This
collisional residue is then blown out of the disk gap by various processes
(see Section \ref{secshock})
and eventually lands in the disk atmosphere where it is observed.

If the colliding rocky objects originated from BP Psc's giant 
star disk material, then they likely formed through similar pathways as planetesimals 
forming around young stars. Another common indicator of planetesimal
formation in young stars is grain growth to mm-cm sizes 
\citep[e.g.,][and references therein]{natta07}.
Is there any evidence for grain growth in the disk around BP Psc? 
Z08 report 880 $\mu$m measurements of
BP Psc and interpret the continuum flux as originating from a population of cooler
or potentially larger grains. The only other longer wavelength data point
that exists for BP Psc that could be used to constrain grain growth
is an upper limit reported at 1 mm (Z08). This upper limit, however,
does not allow one to place any stringent limits on the maximum
grain size within BP Psc's dusty disk.
Limits on the growth of grains in the disk around BP Psc 
will require more constraining observations at wavelengths 
longer than 880 $\mu$m.

There are some pit-falls in this proposed dust enrichment model. 
If the colliding planetesimals are left over from BP Psc's
initial planetary system, then
there must be a large population of trojan-like bodies whose orbits
evolve to bring them closer to the giant planet. In theory, interactions
between such rocky bodies and BP Psc's gaseous disk 
could alter the rocky bodies' orbits and
bring a sufficient number of them close enough to the giant planet to experience
enhanced collisional activity (the likes of which must produce $\sim$1 Lunar mass
of small crystalline grains; see Table 1). 
Of course, if the planetesimals are instead being formed in-situ
in the giant star disk, then this problem disappears.
Another problem is that \citet{bouwman03} do not provide a convincing
argument as to why one should expect isolated regions of nearly 100\%
crystalline dust in the disk atmosphere when the colliding
bodies that produce the dust are more compositionally diverse
\citep[see the middle panel of Figure 5, Figure 10, and Table 3 in][]{bouwman03}.

\subsubsection{Giant Planet Disk Shocking}
\label{secshock}

Another possibility is that the giant planet generates shocks in the disk that 
sufficiently increase the disk temperature to anneal amorphous grains.
Many theoretical works exist that predict shocks in the edges of a disk gap
as a result of tidal interactions between the (gap-opening) giant planet and the disk
\citep[e.g.,][]{takeuchi96,lubow99,rafikov02}. These shocks could act
to heat small, amorphous grains to sufficient temperatures such that
thermal annealing into crystalline species would result \citep{harker02}. The most
interesting aspect of this model is the dust mineralogy it predicts. Annealing
Mg-rich amorphous grains results in the generation of forsterite and 
SiO$_2$ as a by-product,
while reaction of forsterite and SiO$_2$ results in the production
of enstatite \citep[][and references therein]{bouwman01}. Thus,
giant-planet induced shocks in the disk are capable of reproducing
the observed small grain mineralogy. Furthermore, shocks are the
favored production method for cristobalite \citep{sargent09a}, the
most likely polymorph of SiO$_2$ observed in BP Psc's disk.

After shocks produce the crystalline material in BP Psc's disk, the
processed dust must
be relocated from the disk mid-plane into the disk atmosphere
where it is observed. Assuming shock-generated crystalline
material enters the disk gap, it may then be orbitally
perturbed (both radially and vertically) by the giant planet 
such that it may be directly irradiated
by stellar light and driven out of the disk mid-plane by 
radiation pressure. The flared disk geometry will allow
these liberated grains to be re-captured by the disk, although
at larger separations than where they originated. Thus,
the disk atmosphere will become polluted with highly
crystalline material as is observed for BP Psc.

This is our preferred model as it naturally accounts for
both BP Psc's disk structure and mineralogy.

\section{Conclusions}

We have obtained and analyzed Spitzer IRS spectra for the
enigmatic dust and gas-enshrouded star, BP Psc. 
Disk structure analysis suggests that BP Psc's disk is unlike
any classical T Tauri star disk, but is instead similar
to disks around Herbig Ae/Be stars. Such a result suggests that BP Psc
is itself as luminous as these pre-main sequence, intermediate
mass stars. The suite of evidence presented in Z08 requires that
BP Psc $-$ if it is indeed as luminous as a typical HAeBe star $-$ be
a first-ascent giant star.
The match of BP Psc's disk structure with that of the HAeBe star
HD 100546 suggests that BP Psc hosts a gapped disk. Evidence
for a giant planet as the generator of the gap in HD 100546's disk
suggests that BP Psc could also have a giant planet orbiting within its disk.
Dust model fits to the IRS
spectrum show that BP Psc's flared disk atmosphere contains 
chemically and thermally processed, cool, small dust grains.
Our preferred model to explain the location and mineralogy of these
grains relies on a giant planet orbiting BP Psc. 
Disk shocks that result from disk-planet interaction
generate the highly crystalline dust which is subsequently removed
from the disk mid-plane and relocated into the disk atmosphere 
where it is observed. We briefly note, 
as suggested by \citet{bouwman03}, that this highly crystalline dust
could be the seed material that forms cometary bodies like Hale-Bopp.
If this is the case, and if BP Psc is indeed a first-ascent giant star,
then we could be witnessing the initial formation stages
of a second generation of cometary bodies around BP Psc.



\acknowledgments

C.M. acknowledges support from the Spitzer Visiting Graduate Student Program
and from a LLNL Minigrant to UCLA.
We thank the 
IRS GTO team for carrying out these observations and Michiel Min for providing 
us with absorption coefficients for the various dust species. We also thank Carey 
Lisse for useful discussion and providing absorption coefficients for cristobalite. We
thank the anonymous referee for a thorough report that helped improved this paper.
This research was supported in part by NASA grants to UCLA and University of
Georgia.



{\it Facilities:} \facility{Spitzer (IRS)}, \facility{Spitzer (MIPS)}




\clearpage

\clearpage

\clearpage

\clearpage

\clearpage

\begin{deluxetable}{lcccc}
\tabletypesize{\normalsize}
\tablecolumns{5} 
\tablewidth{0pt} 
\tablecaption{Dust Component Parameters \label{tabbpdust}}
\tablehead{ \colhead{Dust Component} &
                      \colhead{Stoichiometry} &
                      \colhead{Temperature} &
                      \colhead{Mass Fraction\tablenotemark{a}} &
                      \colhead{Mass\tablenotemark{b}} \\
                      \colhead{} &
                      \colhead{} &
                      \colhead{(K)} &
                      \colhead{(\%)} &
                      \colhead{(Lunar Masses)}
}
\startdata
Forsterite & Mg$_2$SiO$_4$\tablenotemark{c} & 75$\pm$25 & 59.4$\mathop{}_{-2.4}^{+2.6}$ & 1.2 \\
Enstatite & MgSiO$_3$\tablenotemark{c} & 75$\pm$25 & 36.2$\mathop{}_{-2.8}^{+2.4}$ & 0.6 \\
Olivine    & Mg$_2$SiO$_4$\tablenotemark{d} & 150$\pm$25 & 0.06$\mathop{}_{-0.06}^{+0.35}$ & $-$\tablenotemark{e} \\
Pyroxene & MgSiO$_3$\tablenotemark{d} & 150$\pm$25 & 4.0$\mathop{}_{-0.5}^{+0.3}$ & 6$\times$10$^{-2}$ \\
Cristobalite & SiO$_2$\tablenotemark{c} & 250$\pm$50 & 0.03$\mathop{}_{-0.01}^{+0.06}$ & 6$\times$10$^{-4}$ \\
Continuum 1 & $-$ & 1000$\mathop{}_{-100}^{+50}$ & $-$   & \multirow{3}{*}{600\tablenotemark{f}} \\
Continuum 2 & $-$ & 250$\mathop{}_{-35}^{+25}$   & $-$   & \\
Continuum 3 & $-$ & 175$\mathop{}_{-30}^{+25}$  & $-$  & \\
\enddata
\tablenotetext{a}{Fraction of mass in 0.1-1.0 $\mu$m, solid-state emitting grains detected in our IRS spectrum 
                               only. The reported percentage does not
                               include the mass contained in the material responsible for the 
                               continuum emission.}
\tablenotetext{b}{Assuming the mass of the moon is 7.35$\times$10$^{25}$ g and
                               that BP Psc is a giant star 300 pc distant from the Earth (Z08). 
                               For an assumed distance
                               of 80 pc, appropriate if BP Psc were a T Tauri star, 
                               the mass per grain species is reduced by a factor of $\sim$15.}
\tablenotetext{c}{Crystalline grains (see Section \ref{secbppscirsdust}).}
\tablenotetext{d}{Amorphous grains (see Section \ref{secbppscirsdust}).}
\tablenotetext{e}{We do not consider amorphous olivine as significantly detected in our
                               final $\chi$$^2$ model fit.}
\tablenotetext{f}{This mass
                               is from Z08 where they computed the disk dust mass
                               assuming half of the 880 $\mu$m detected flux towards BP Psc comes from
                               a region of the disk having T$_{\rm dust}$= 36 K
                               \citep[fits to our IRS spectrum do not require that an additional $\sim$36 K dust 
                               region exists; Figure 4 of][shows how the 33 $\mu$m forsterite feature
                               is sensitive to cool material]{malfait98}. If there is no 36 K dust, then the
                               mass listed for the continuum component could be a 
                               few times smaller.}
\end{deluxetable}

\end{document}